\documentclass[twocolumn,showpacs,preprintnumbers,amsmath,amssymb]{revtex4}

\usepackage{amsmath}
\usepackage{amsfonts}
\usepackage{amssymb, multirow}
\usepackage{pdfsync}
\usepackage{epsfig}
\usepackage{graphicx}
\usepackage{nicefrac}

\newcommand{\be}{\begin{equation}}\newcommand{\ee}{\end{equation}}
\newcommand{\bea}{\begin{eqnarray}} \newcommand{\eea}{\end{eqnarray}}
\newcommand{\ba}[1]{\begin{array}{#1}} \newcommand{\ea}{\end{array}}

\long\def\symbolfootnote[#1]#2{\begingroup%
\def\thefootnote{\fnsymbol{footnote}}\footnote[#1]{#2}\endgroup} 

\newcommand{\cN}{{\cal N}}

\def\bfone{\relax{\rm 1\kern-.35em 1}}

\begin{document}


\title{Triality, Periodicity and Stability of $\textrm{SO}(8)$ Gauged Supergravity}

\author{Andrea Borghese$^1$, Adolfo Guarino$^2$ and Diederik Roest$^1$}

\affiliation{%
~ \\
$^1$Centre for Theoretical Physics, University of Groningen, Nijenborgh 4 9747 AG Groningen, The Netherlands
\\
\\
$^2$Albert Einstein Center for Fundamental Physics, Institute for Theoretical Physics, Bern University, Sidlerstrasse 5, CH–3012 Bern, Switzerland
}%

\begin{abstract}

While electromagnetic duality is a symmetry of many supergravity theories, this is not the case for the $\cN = 8$ gauged theory. It was recently shown that this rotation leads to a one-parameter family of $\textrm{SO}(8)$ supergravities. It is an open question what the period of this parameter is. This issue is investigated in the $\textrm{SO}(4)$ invariant sectors of the theory. We classify such critical points and find a {\it novel} branch of non-supersymmetric and unstable solutions, whose embedding is related via triality to the two known ones. Secondly, we show that the three branches of solutions lead to a $\pi/4$ periodicity of the vacuum structure. The general interrelations between triality and periodicity are discussed. Finally, we comment on the connection to other gauge groups as well as the possibility to achieve (non-)perturbative stability around AdS/Mkw/dS transitions.

\end{abstract}

\maketitle

\section{Introduction}

The equivalence of the electric and magnetic formulation of Maxwell's equations lies at the heart of our understanding of electromagnetism. In the absence of electric sources, it is fundamentally impossible to distinguish  between the dual pictures. The theory is indifferent as to whether the $\textrm{U}(1)$ gauge vector is electric, magnetic or any linear, dyonic combination thereof.

A similar phenomenon is at work in most supersymmetric theories. The duality symmetry of such theories comprises the R-symmetry group, which in four dimensions generically is $\textrm{SU}(\cN) \times \textrm{U}(1)$ for $\cN$-extended supersymmetry. The latter factor of this symmetry rotates electric and magnetic vectors into each other \footnote{Another example is $\cN = 4$ super-Yang-Mills theory; in this case, electromagnetic duality is included in the $\textrm{SL}(2,\mathbb{R})$ duality symmetry of this theory.}.

In maximal supergravity, however, the R-symmetry is $\textrm{SU}(8)$ and lacks the $\textrm{U}(1)$ factor. As a consequence, electromagnetic duality is {\it not} a part of the $\textrm{E}_{7(7)}$ duality symmetry of the $\cN = 8$ theory. Instead, it is only included in the $\textrm{Sp}(56,\mathbb{R})$ group, spanning all transformations between the 28 electric and 28 magnetic gauge vectors. These will generically not be symmetries of the theory, but relate different symplectic frames. However, as these stem from an (albeit non-local) field redefinition, physics will not be affected by the choice of symplectic frame.

The situation changes radically when one includes charges. The absence of magnetic monopoles breaks the duality symmetry in electromagnetism. A similar mechanism turns out to be at work in maximal supergravity. The introduction of charges in that case is analogous to considering gauged supergravities, where a subgroup of the $\textrm{E}_{7(7)}$ global symmetry group has been gauged. The prime example is the $\textrm{SO}(8)$ theory \cite{Cremmer:1979up,de Wit:1982ig}, whose relevance ranges from Kaluza-Klein theory \cite{de Wit:1986iy} to holography \cite{Aharony:2008ug}. For almost three decades after its inception, this theory was implicitly assumed to be unique. Very recently it was realised, however, that different theories are obtained depending on whether the starting point includes electric, magnetic or dyonic gauge vectors \cite{Dall'Agata:2012bb}. As a consequence, there is a one-parameter family of $\textrm{SO}(8)$ theories with different physical properties. It will be denoted by $\omega$, where $\omega = 0$ ($\omega = \pi/2$) corresponds to electric (magnetic) vectors.

A natural question concerns the periodicity of the new parameter. Based on a group-theoretic argument involving $\textrm{E}_{7(7)}$ invariants, the period was shown to be no less than $\pi/4$ \cite{Dall'Agata:2012bb}. As a result, it was conjectured that at the intermediate value $\omega = \pi/8$, $\textrm{SO}(8)$ supergravity should be identified as the gravity dual of a particular $D=3$ SCFT \cite{ABJ}. Similar considerations will apply to the $\omega$-range when a higher-dimensional origin for the one-parameter family is proposed. It is therefore of importance to investigate the periodicity of $\omega$ from various points of view and see whether these are consistent with $\pi/4$ or a multiple thereof.

An example of such a line of research is provided by the vacuum analysis of the $\textrm{G}_2$ and $\textrm{SU}(3)$ invariant sector of the full theory \cite{Dall'Agata:2012bb,Borghese:2012qm,Borghese:2012zs}. Indeed, it was found that the set of all vacua is also periodic in $\pi/4$, consistent with the earlier conjecture. It will be instructive for what follows to recall these results for the special case of $\textrm{SO}(7)$ invariant vacua.

The standard formulation of maximal supergravity has two critical points that preserve an $\textrm{SO}(7)_-$ subgroup of $\textrm{SO}(8)$, while for the $\textrm{SO}(7)_+$  embedding one finds just a single point \cite{Warner:1983vz}. While the mass spectra are identical, the two sets differ on their values of the scalar potential $V$. Turning on the new parameter, one finds that both the location of the points and the value $V$ are $\omega$-dependent while the mass spectrum is independent. Interestingly, one of the two $\textrm{SO}(7)_-$ critical points moves to the boundary of moduli space at $\omega = \pi/4$, while the other takes on all properties of the old $\textrm{SO(7)}_+$ point, including the value of $V$. 
Indeed, the $\textrm{SO(7)}_-$ sector is identical to that of $\textrm{SO(7)}_+$ after a $\pi/4$ shift of $\omega$. As a result, the combination of both sectors is $\pi/4$ periodic.

An important test of the conjectured periodicity consists of the $\textrm{SO}(4)$ invariant critical points. Similar to the $\textrm{SO}(7)$ sector, two types of such points are known; however, in contrast to it, these differ both in the value of the scalar potential as well as in the normalised mass spectrum. One such branch has been known since the 1980s; however, only recently its full mass spectrum was calculated and found to be perturbatively stable \cite{Warner:1983du,Fischbacher:2010ec}. Moreover, an unstable second branch was found with an inequivalently embedded residual gauge group \cite{Fischbacher:2010ec}. We will refer to these embeddings as $\textrm{SO}(4)_{v,s}$, respectively, where the subscripts refer to the three inequivalent $\bf 8$-irreps of $\textrm{SO}(8)$ that are a consequence of its triality. This strongly suggests a third embedding, $\textrm{SO}(4)_c$. Indeed, in section II we will show that there is {\it a new critical point} with this residual gauge group and a distinct mass spectrum.

In section III we will analyse how the three critical points are related when turning on the parameter $\omega$, and investigate which of the three critical points and their mass spectra morph into each others as a function of $\omega$. It will turn out that {\it the points with $\textrm{SO}(4)_{s,c}$ are related} in exactly the same way as $\textrm{SO}(7)_\pm$ by $\pi/4$ shifts of the parameter $\omega$. In this way the appearance of the new, triality related solutions indeed allows the periodicity of the $\textrm{SO}(8)$ vacuum structure to be $\pi/4$.

In order to derive the full mass spectra of these solutions, we change gears in section IV and explore $\textrm{SO}(4)$ invariant critical points of other theories. This will naturally lead to connections between the $\textrm{SO}(8)$ supergravity and other theories with non-compact gaugings. As a result, we will connect the unstable AdS solution of the $\textrm{SO}(8)$ theory in refs~\cite{Warner:1983du,Fischbacher:2010ec} to the novel dS solutions of the $\textrm{SO}(4,4)$ theory in ref.~\cite{Dall'Agata:2012sx}. We will discuss how this relation might help to achieve {\it \text{(non-)perturbative} stability} close to Minkowski vacua.

\section{Triality of $\textrm{SO}(4)$-sectors}

Maximal supergravity contains $70$ physical scalars $\phi_{IJKL}=\phi_{[IJKL]}$ satisfying the self-duality (SD) condition
\be
\label{vD_condition}
\phi_{IJKL} = \frac{1}{4!} \, \epsilon_{IJKLMNPQ} \, \phi^{MNPQ} \ ,
\ee
with $\phi^{IJKL}=(\phi_{IJKL})^{*}$, where $I=1,..,8$ refers to the fundamental representation of the (local) R-symmetry group $\textrm{SU}(8)$. These serve as coordinates in a coset space $\textrm{E}_{7(7)}/\textrm{SU}(8)$. We identify the $\bf 8$ of $\textrm{SU}(8)$ with the ${\bf 8}_v$ of the gauge group $\textrm{SO}(8)$ without loss of generality. Decomposing the scalar fields \eqref{vD_condition} leads to ${\bf 35}_s$ and ${\bf 35}_c$ irreps, which are (anti-)self dual four-forms in terms of the vectorial index. These are the real and imaginary parts of $\phi_{IJKL}$.

In order to discuss the issue of periodicity of electromagnetic rotations, we will restrict to a consistent truncation to a sector of the theory that retains only the invariant scalars under an $\textrm{SO}(4)$ subgroup of the R-symmetry group. However, the embedding ${\textrm{SO}(4) \subset \textrm{SO}(8)}$ is not unique due to $\textrm{SO}(8)$ triality. For this reason we will consider three scalar sectors arising from inequivalent embeddings, which we will denote by $\textrm{SO}(4)_{v,s,c}$. In what follows we will discuss these three consecutively. In this section we restrict ourselves to the standard formulation of maximal supergravity ($\omega = 0$).

\subsection{The vectorial embedding}

The $\textrm{SO}(4)_v$ embedding leads to the decomposition
 \begin{align}
\bold{8}_{v} \rightarrow  \bold{1} + \bold{3}_{1} + \bold{1} +\bold{3}_{2} \,, \quad \bold{8}_{s,c} \rightarrow \bold{4} + \bold{4} \ .
\end{align}
This induces an index splitting for the $\bold{8}$ of $\textrm{SU}(8)$ of the form $I=1 \oplus a \oplus \hat{1} \oplus \hat{a}$ with $a=2,3,4$ and $\hat{a}=\hat{2},\hat{3},\hat{4}$. Therefore, only two invariant complex scalars are retained
\be
\label{scalarsBos}
\phi_{1 abc} = (\phi_{\hat{1} \hat{a}\hat{b}\hat{c}})^{*} = z_{1}  \, \epsilon_{abc} \,\,\, , \,\,\,
\phi_{\hat{1} abc} =  - (\phi_{1 \hat{a}\hat{b}\hat{c}})^{*} = z_{2} \, \epsilon_{abc} ,
\ee
with $z_{1} \equiv\rho_1 \, e^{i \alpha_1}$ and $z_{2} \equiv \rho_2 \, e^{i \alpha_2}$. 

The $\textrm{SO}(4)_v$ invariant scalar potential (we set $g=1$) for the one-parameter family of $\textrm{SO}(8)$ gaugings is (the subscript refers to the value of $\omega$)
\bea
\label{s_Bos}
 V^v_{0} &=& -\dfrac{1}{8 \rho^4} \Big(   32 \rho_{1}^4+61 \rho_{1}^2 \rho_{2}^2+32 \rho_{2}^4 \nonumber \\[1mm]
                            &+& 4 \cosh (2 \rho ) \left(4 \rho_{1}^4+9 \rho_{1}^2 \rho_{2}^2+4 \rho_{2}^4\right) \\[1mm]
                            &-&  \rho_{1}^2 \rho_{2}^2 \left(\cosh (4 \rho )-8 \cos (2 \alpha ) \sinh ^4(\rho )\right) \Big) \ ,\nonumber
\eea
where $\rho^2\equiv\rho_1^2 + \rho_2^2$ and $\alpha\equiv\alpha_1-\alpha_2$, while  the orthogonal combination $\alpha_{\perp}\equiv\alpha_1+\alpha_2$ does not enter the potential. 

\begin{figure}[ht]
\includegraphics[width=50mm]{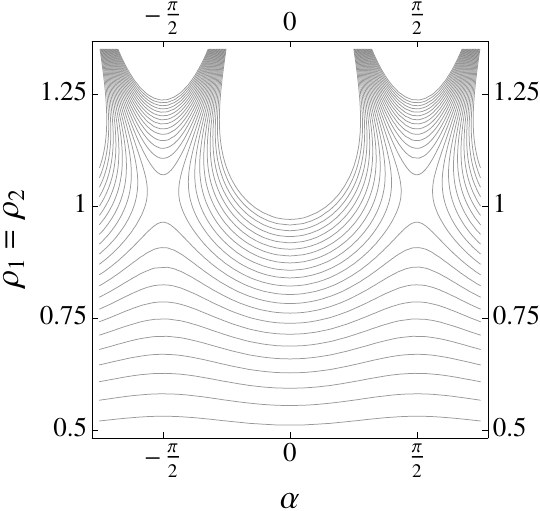}
\caption{{\it Contours of the scalar potential $V_0^v$ for $\rho_{1}=\rho_{2}$.}}
\label{Fig:Bos_critical_points} 
\end{figure}

The extremisation condition $\partial_{\alpha} V_0^v=0$ imposes ${\alpha=n \, \frac{\pi}{2}}$ for $n = 0, \pm 1 , 2$. If $n$ is even, then the only critical point of the scalar potential is located at the origin $\rho_{1}=\rho_{2}=0$ with $V_0^v=-6$ and the normalised masses are given by $m^2 L^2=-2 \,(\times 4)$. It corresponds to the $\textrm{SO}(8)$ invariant extremum preserving $\mathcal{N}=8$ supersymmetry. For odd values of $n$, there are two extrema with ${\alpha=\pm \frac{\pi}{2}}$ and $\rho_1=\rho_{2}=\frac{1}{\sqrt{2}} \textrm{cosh}^{-1}(\sqrt{5})$ (see FIG.~\ref{Fig:Bos_critical_points}). The energy at the two extrema turns out to be $V_0^v=-14$ and the normalised masses are $m^2 L^2=60/7, -12/7  \, (\times 2), 0  $. These solutions were originally found in ref.~\cite{Warner:1983du} while their full mass spectra were computed more recently in ref.~\cite{Fischbacher:2010ec} and found to be
\be
\begin{array}{ccl}
m^2 L^2 &=&  \tfrac{3}{7} (5 \pm \sqrt{65}) \,\, (\times 9) \,\, , \,\, \tfrac{18}{7} \,\, (\times 9)  \\[2mm]
               & &  \tfrac{60}{7} \,\, (\times 1) \,\, , \,\,  -\tfrac{12}{7} \,\, (\times 20) \,\, , \,\,   0 \,\, (\times 22) \ .
\end{array}
\ee
They are non-supersymmetric and nevertheless satisfy the BF bound $m^2 L^2 \geq  -9/4$ for stability in AdS.

\subsection{The spinorial embedding}

The $\textrm{SO}(4)_s$ embedding leads to the decomposition
\begin{align}
\label{8_spinorial}
\bold{8}_{s} \rightarrow  \bold{1} + \bold{3}_{1} + \bold{1} +\bold{3}_{2} \,, \quad \bold{8}_{c,v} \rightarrow \bold{4} + \bold{4} \ .
\end{align}
In this case the index splitting for the $\bold{8}$ of $\textrm{SU}(8)$ reads $I=i \oplus \hat{i}$ with $i=1,2,3,4$ and $\hat{i}=\hat{1},\hat{2},\hat{3},\hat{4}$. The two indices $i$ and $\hat{i}$ transform in (two copies of) the vector representation. The SD condition in (\ref{vD_condition}) is satisfied by the $\textrm{SO}(4)_s$ invariant scalars 
\bea
\label{scalarsFermi_common}
\phi_{ijkl} &=& \phantom{-} (\phi_{\hat{i}\hat{j}\hat{k}\hat{l}})^{*}  \,\,=\,\,  \phantom{\tfrac{1}{2}} \, (x_{1} + i y_{1}) \, \epsilon_{ijkl} \  , \nonumber \\
\phi_{\hat{i}jkl} &=& - (\phi_{i\hat{j}\hat{k}\hat{l}})^{*}  \,\,=\,\,  \tfrac{1}{2} \, (x_{2} + i y_{2}) \, \epsilon_{\hat{i}jkl} \ , \\[1mm]
\phi_{i\hat{j} k\hat{l}}  & = &   x_3 \, \epsilon_{i\hat{j} k\hat{l}}  \,\, + \,\, x_4  \, \delta_{[i\hat{j} } \delta_{k\hat{l}]} \nonumber \ ,  
\eea
with $x_{1,2,3,4} \in \bold{35}_{s}$ and ${y_{1,2} \in \bold{35}_{c}}$. In order to get a managable scalar potential we mod out the theory by a discrete $D_{4}$ subgroup of $\textrm{SU}(8)$ that is an automorphism of $\textrm{SO}(4)_s$. It is generated by \footnote{We are aware of at least one different embedding of $D_4$ that leads to the same two-scalar potential. Hence there is a multiple of the corresponding critical points.}
\be
\label{D4_subgroup}
S_{0}=\left(
\begin{array}{cc}
\mathbb{I}_{4} & 0 \\
0 & -\mathbb{I}_{4}
\end{array}\right)
\hspace{2mm} , \hspace{2mm}
S_{1}=\left(
\begin{array}{cc}
0 & \eta \\
\eta & 0
\end{array}\right) \ ,
\ee
with $\eta=\textrm{diag}(1,-1,-1,-1)$. This projects out four of the scalars keeping only $(x_{4},y_{1})$. The scalar potential for these two fields is 
\be
\begin{array}{ccl}
\label{s_Fermi}
V^s_{0}  & = & \tfrac12 \sinh ^2(y_{1}) \left(\cosh (6 x_{4}) - 4 \sinh ^3(2 x_{4})\right)   \\[2mm]
                        &-& \tfrac34  \cosh (2 x_{4}) (3 \cosh (2 y_{1})+5) \ .
\end{array}
\ee
It coincides with that of ref.~\cite{Fischbacher:2010ec} after the field redefinitions ${x_{4}=-\log(\rho)/2}$ and $y_{1}=\lambda$.

\begin{figure}[t]
\includegraphics[width=50mm]{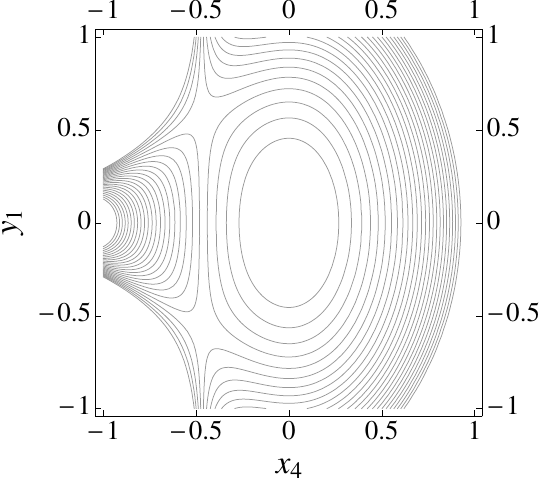}
\caption{{\it Contours of the scalar potential $V^s_0$.}}
\label{Fig:Ferm_critical_points_v} 
\end{figure}

Turning to the critical points of this potential (see FIG.~\ref{Fig:Ferm_critical_points_v}), there are two critical points in addition to the maximally symmetric one \cite{Fischbacher:2010ec}. They form a conjugate pair of non-supersymmetric critical points located at
 \begin{align} \label{newWarner}
  (x_4,y_{1})=(\log [2/\sqrt{3}-1]/4 \,,\, \pm \cosh ^{-1}[\sqrt{3}]/2) \ ,
 \end{align}
and have an energy 
 \begin{align}
 V_0^s=-2\sqrt{9+6\sqrt{3}} \ .
 \end{align}
The normalised masses for the two  scalars at these solutions are $m^2 L^2= 4 \sqrt{3} , -2 \sqrt{3}$. The full mass spectrum is given by
\be
\label{masses_new_Warner}
\begin{array}{ccl}
m^2 L^2 &=&   3+\sqrt{3}\pm\sqrt{6 (4+\sqrt{3})} \,\, (\times 1)  \\[2mm]
&&\tfrac{3}{2} (\sqrt{3}-3) \,\, (\times 15)  \,\, , \,\, 2 (\sqrt{3}-3) \,\, (\times 10)  \\[2mm]
&&2 (\sqrt{3}-2) \,\, (\times 9) \,\, , \,\, 2 (\sqrt{3}-1) \,\, (\times 9)  \,\, , \,\, 0 \,\, (\times 22)  \\[2mm]
&&\tfrac{3}{2} (5+\sqrt{3} ) \,\, (\times 1) \,\, , \,\, 4 \sqrt{3} \,\, (\times 1) \ , -2 \sqrt{3} \,\, (\times 1)  \ ,
\end{array}
\ee
showing two unstable masses that violate the BF bound.

\subsection{The conjugate embedding}

The third inequivalent embedding $\textrm{SO}(4)_c$ implies
\begin{align}
\label{8_conjugate}
\bold{8}_{c} \rightarrow  \bold{1} + \bold{3}_{1} + \bold{1} +\bold{3}_{2} \,, \quad \bold{8}_{v,s} \rightarrow \bold{4} + \bold{4} \ .
\end{align}
The index splitting $I=i \oplus \hat{i}$ therefore coincides with that of the $\textrm{SO}(4)_{s}$ sector and hence also the set of invariant tensors. The SD condition in (\ref{vD_condition}) is now satisfied by the real scalars in the first two lines of \eqref{scalarsFermi_common} augmented with
\be
\label{scalarsFermi_differentSO4c}
\begin{array}{cccclcccc}
\phi_{i\hat{j} k\hat{l}} &=&   i \,  y_3 \, \epsilon_{i\hat{j}k\hat{l}} \,\, + \,\, i  \, y_4 \, \delta_{[i \hat{j}} \, \delta_{k \hat{l}]}  & ,  
\end{array}
\ee
with $x_{1,2} \in \bold{35}_{s}$ and $y_{1,2,3,4} \in \bold{35}_{c}$. Modding out again by a discrete $D_{4}$ subgroup as in (\ref{D4_subgroup}) but now with ${\eta=\textrm{diag}(1,1,1,1)}$,  retains $(y_{4},x_{1})$. 

\begin{figure}[t]
\includegraphics[width=50mm]{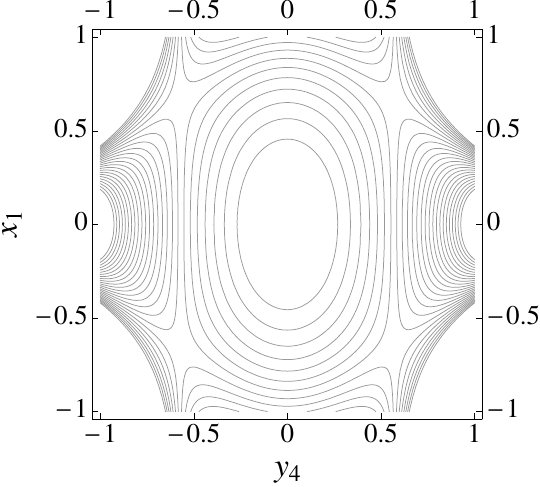}
\caption{{\it Contours of the scalar potential $V_0^c$.}}
\label{Fig:Ferm_critical_points_s} 
\vspace{-1cm}
\end{figure}

The scalar potential for this new sector is computed to be 
\be
\begin{array}{ccl}
\label{s_FermiSO4c}
V_0^c  &=&   \tfrac12 \sinh ^2(x_{1}) \cosh (6 y_{4})   \\[2mm]
                  &-&  \tfrac34 \,  \cosh (2 y_{4}) \, (3 \cosh (2 x_{1})+5)   \ . 
\end{array}
\ee
It can be seen from FIG.~\ref{Fig:Ferm_critical_points_s} that this  sector has four new critical points. They turn out to break all supersymmetry and to be located at
 \begin{align}
 \label{location_new_solutions}
  (y_{4},x_{1})=(&\pm \cosh ^{-1}[ (1/2+\sqrt{3}/2)^{1/2}] \,\, ,  \, \notag \\
 &\pm \cosh ^{-1}[\sqrt{3}/\sqrt{2}]) \ ,
 \end{align}
with unrelated sign choices, and have an energy 
\be
\label{s_Pi/4}
V_0^c=-6 \, \sqrt{3} \ .
\ee
The two scalar masses are $m^2 L^2 = 2 \, (1 \pm \sqrt{7})$ while the full scalar mass matrix is given by
\be
\label{masses_new_solutions}
\begin{array}{ccl}
m^2 L^2 &=&   2(1 \pm \sqrt{7}) \,\, (\times 1)  \,\, , \,\, -2  \,\, (\times 10)  \\[2mm]
&&  5\pm\sqrt{37} \,\, (\times 1)  \,\, , \,\, 1\pm\sqrt{5} \,\, (\times 9) \\[2mm]
&& -\tfrac{3}{2}  \,\, (\times 15)  \,\, , \,\, \tfrac{21}{2} \,\, (\times 1) \,\, , \,\, 0 \,\, (\times 22)  \ ,
\end{array}
\ee
as will be derived in section IV. This solution turns out to be unstable. Notice that the trace of the mass matrix vanishes in this critical point.

At first sight it may seem surprising that this set of points seems to have been missed in the numerical scan of ref.~\cite{Fischbacher:2011jx}, as their vacuum energy lies in the investigated range of that reference. However, these numerical methods have trouble reproducing some of the more symmetric critical point: for instance, the algorithm could not retrieve the $\textrm{SO}(4)_s$ vacuum by its own. It could thus well be that a similar caveat applies to the new $\textrm{SO}(4)_c$ point found here.

\section{Periodicity of vacuum structure}

In this section we leave the standard formulation of maximal supergravity and consider so-called new maximal supergravity, where an arbitrary combination of electric and magnetic vectors serve as gauge vectors. Our focal point will be the periodicity of the full set of $\textrm{SO}(4)$ invariant critical points. To this end we will construct the generalisation of the scalar potentials of the previous section by including the electromagnetic phase $\omega$.

An explicit calculation demonstrates that the \text{$\omega$-parameter} drops out of the scalar potential for the vectorial embedding
\begin{align}
 V^v = V^v_{\omega = 0} \ .
 \end{align}
This remarkable result, which has no counterpart in any other $\textrm{SO}(4)_{s,c}$ or $\textrm{SU}(3)$ invariant sector, implies that both the location, the cosmological constant, as well as the mass spectra of the $\textrm{SO}(4)_v$ points are the same for all values of $\omega$.

The situation is yet more interesting when turning to the spinorial $\textrm{SO}(4)_{s}$ embedding. The full, $\omega$-dependent scalar potential for the two fields is calculated to be 
\be
\label{s_Fermi}
V^s  =  \cos^2(\omega)  \,\,  V^s_{0} (x_4,y_{1})  \,+\, \sin^2(\omega)  \,\,  V^s_{0} (-x_4,y_{1})  \ .
\ee
Note that the scalar potential is invariant under a $\pi /2$ shift of $\omega$ accompanied by a sign flip of $x_4$. This sector by itself therefore has a $\pi/2$ periodicity. Moreover, the potential does not depend on the sign of $\omega$ and as a consequence we only have to consider a $\pi/4$ parameter range. 

For an intermediate value $0<\omega \leq \pi/4$ specifying a dyonic gauging, there are four critical points in addition to the maximally symmetric one. They split into two pairs of degenerate solutions, each of which produces a  different scalar mass spectrum and energy. When $\omega$ approaches the boundary values $\omega \rightarrow 0$, one of the pairs of critical points runs away in field space with asymptotic energy $V \rightarrow -\infty$ and scalar masses $m^2 L^2 \rightarrow 3 \pm \sqrt{33}$. Moreover, at the special value $ \omega =\pi/4$, the four critical points become degenerate in energy and scalar mass spectra.  

Finally, the $\omega$-dependent scalar potential of the restricted two-field model for the conjugate $\textrm{SO}(4)_{c}$ embedding is given by
\be
\label{s_FermiSO4c}
V^c  =  V^c_0 \,+ 4 \, \sin{\omega} \cos{\omega} \, \sinh ^2(x_1) \sinh ^3(2 y_4)  \ .
\ee
Remarkably, this scalar potential is exactly the same as that of the $\textrm{SO}(4)_s$ sector provided one shifts $\omega$ by $\pi/4$ and replaces $x_{4} \leftrightarrow y_{4}$ and $y_1 \leftrightarrow x_{1}$. Thus, with this third branch of $\textrm{SO}(4)$ critical points, the periodicity of the full $\textrm{SO}(4)$ vacuum structure boils down to $\pi/4$. 

Note the distinct pattern under $\omega$-rotations: of the three triality-related sectors, the vectorial one remains invariant while the spinorial and conjugate are rotated into each other. The $\pi/4$ periodicity crucially relies on the existence of these inequivalent embeddings under triality. 

The same behaviour can be seen for the $\textrm{SO}(7)$ sectors. There are also three triality-related embeddings in $\textrm{SO}(8)$, of which $\textrm{SO}(7)_\pm$ are rotated into each other under $\pi/4$ shifts of the electromagnetic parameter $\omega$ \cite{Borghese:2012zs}. In this case, however, the third embedding $\textrm{SO}(7)_v$ leads to an empty sector: no scalar is invariant and hence there are no critical points.

\section{Singular limits and stability}

For the $\textrm{SO}(4)_{s,c}$ embeddings, we showed that the scalar potential possesses two pairs of critical points one of which runs away  when $\omega$ approaches specific values, for instance $\omega \rightarrow 0$ in the spinorial embedding. It is natural to wonder what happens to these solutions in such singular limits. In order to address this question, we will move to an alternative approach  known as the ``go to the origin" (GTTO) approach. It was introduced in the context of half-maximal supergravity \cite{Dibitetto:2011gm} and later applied to maximal supergravity \cite{DallAgata:2011aa,Borghese:2012qm,Kodama:2012hu,Borghese:2012zs} very successfully. The main idea underlying the GTTO approach is to look for theories compatible with a given critical point instead of looking for critical points compatible with a given theory. We will  introduce it only briefly; further details can be found in the references quoted above. 

The GTTO approach makes use of the formulation of maximal supergravity in terms of the fermion mass terms $\mathcal{A}^{IJ}(\phi)$ and ${\mathcal{A}_{I}}^{JKL}(\phi)$, which must obey certain linear and quadratic constraints (QC) \cite{de Wit:2007mt}. Once a critical point is found at $\phi=\phi_{0}$ preserving some residual symmetry $G_{res}$, then the fermion masses $\mathcal{A}^{IJ}(\phi_{0})$ and ${\mathcal{A}_{I}}^{JKL}(\phi_{0})$ can be written in terms of invariant tensors of $G_{res}$. Since the scalar manifold in maximal supergravity is a homogeneous space -- all points are connected by $\textrm{E}_{7(7)}$ transformations --, one can always bring any critical point to the origin $\phi_{0}=0$ without loss of generality. The fermion masses will then adopt a form compatible with the residual symmetry. This allows us to identify the non-vanishing components of $\mathcal{A}^{IJ}(0)$ and ${\mathcal{A}_{I}}^{JKL}(0)$, solve the QC and the equations of motion (EOM) of maximal supergravity and find the configurations compatible with a critical point at the origin preserving $G_{res}$. More interestingly, these configurations may depend on some parameters covering the range of gaugings that produce such a critical point and hence following its evolution along different theories. Loosely speaking, what one does is to focus on a critical point and follow it along the different theories (gaugings) that are compatible with it.

\begin{figure}[ht]
\includegraphics[width=70mm]{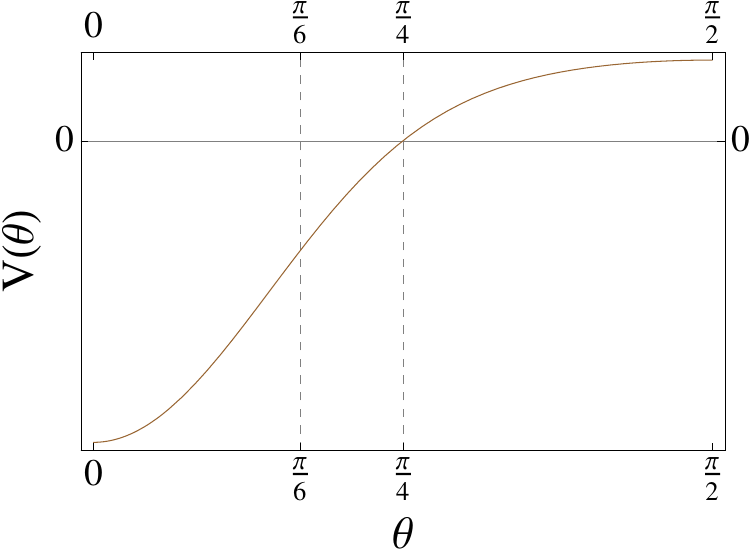}
\includegraphics[width=74mm]{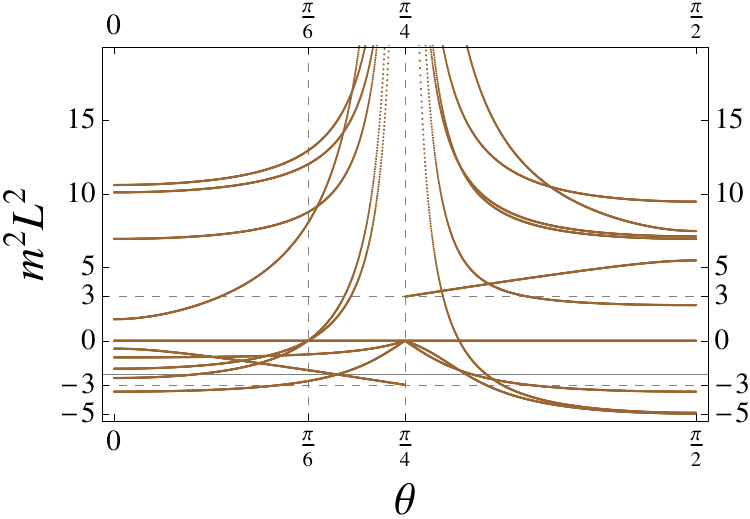}
\caption{{\it Top: Scalar potential interpolating between the AdS solutions at $\theta=0$ and the dS one at $\theta=\frac{\pi}{2}$. The sequence of underlying gaugings reads: \\[1mm]
 - $\textrm{SO}(8)$ for $\theta \in [0,\pi/6)$  , \\[1mm]
 - $\textrm{SO}(2) \times \textrm{SO}(6) \ltimes T^{12}$ at $\theta=\pi/6$  , \\[1mm]
 - $\textrm{SO}(6,2)$ for $\theta \in (\pi/6,\pi/4)$  ,  \\[1mm]
 - $\textrm{SO}(3,1)^2 \ltimes T^{16}$ at ${\theta=\pi/4}$ , \\[1mm]
 - $\textrm{SO}(4,4)$ for $\theta \in (\pi/4,\pi/2]$  .  \\[1mm]
Bottom: Evolution of the scalar masses as a function of the parameter $\theta \in [0,\frac{\pi}{2}]-\{\frac{\pi}{4}\}$. The horizontal solid line denotes the BF bound $m^2 L^2 \geq  -9/4$ for stability in the AdS side.}
}
\label{Fig:AdSTodS} 
\vspace{-.5cm}
\end{figure}

In order to make contact with the previous sections, let us require the residual symmetry to be ${G_{res}=  \textrm{SO}(4)}$ embedded accordingly to the index splitting ${I = i \oplus \hat{i}}$. In addition, we will further mod out by the $D_{4}$ subgroup in (\ref{D4_subgroup}). The components in $\mathcal{A}^{IJ}(0)$ compatible with these symmetries are given by
\be
\label{A1_config}
\mathcal{A}^{ij}= \alpha \,\, \delta^{ij} \hspace{5mm} , \hspace{5mm}  \mathcal{A}^{\hat{i}\hat{j}} = \alpha \,\, \delta^{\hat{i}\hat{j}} \ ,
\ee
whereas those in ${\mathcal{A}_{I}}^{JKL}(0)$ read
\be
\label{A2_config}
\begin{array}{lclc}
{\mathcal{A}_{i}}^{jkl}= \phantom{-} \beta \,\, {\epsilon_{i}}^{jkl} & , & {\mathcal{A}_{i}}^{\hat{j}\hat{k}l}=  - \delta \,\, {\epsilon_{i}}^{\hat{j}\hat{k}l} +  \gamma \,\, \delta_{i}^{[\hat{j}}  \delta^{\hat{k}] l} & ,\\[2mm]
{\mathcal{A}_{\hat{i}}}^{\hat{j}\hat{k}\hat{l}}= - \beta \,\, {\epsilon_{\hat{i}}}^{\hat{j}\hat{k}\hat{l}}
& , &
{\mathcal{A}_{\hat{i}}}^{jk\hat{l}}= \phantom{\mp} \delta  \,\, {\epsilon_{\hat{i}}}^{jk\hat{l}}  +  \gamma \,\,  \delta_{\hat{i}}^{[j}  \delta^{k] \hat{l}} & ,
\end{array}
\ee
with $\alpha , \beta , \delta , \gamma \in \mathbb{C}$. 
%
%
Imposing the set of QC and EOM, we find a one-parameter family of fermion masses \footnote{There is also freedom for an overall scaling parameter $(\mathcal{A}^{IJ},{\mathcal{A}_{I}}^{JKL}) \rightarrow \lambda \, (\mathcal{A}^{IJ},{\mathcal{A}_{I}}^{JKL})$ that we have set to one for the sake of simplicity.}. We will denote the parameter $\theta$. Due to the imposed symmetries, this solution must coincide with the $\omega$-evolution of the critical point \eqref{newWarner} for some range of $\theta$. The fermion mass terms are specified by
\begin{alignat}{2}
& \alpha = A \, e^{i \theta} (\sqrt{2} \sin (2 \theta ) - i B) \,\,\,\,\,\,\, \ ,  && \gamma =-i \, 2\, \sqrt{2} \, A \, e^{-i \theta} \ , \notag \\
& \beta =-e^{-i \theta} ( B/\sqrt{2} - i \sin(2 \theta) ) \,\ , \quad && \delta = e^{i \theta} \,,
\end{alignat}
with the functions $A=\frac{1}{2} ( \sqrt{2} \cos (2 \theta ) B +\cos (4 \theta )+3)^{1/2}$ and ${B=(\cos (4 \theta )+5)^{1/2}}$. The scalar potential can be computed in terms of the fermion masses \cite{de Wit:2007mt}:
\be
\label{sAdSTodS}
V(\theta) = -6 \,\, ( \,1+\cos (4 \theta )+\sqrt{2} \, \cos (2 \theta ) \, B \, ) \ ,
\ee
which has a period of $\pi$ and further exhibits a reflection symmetry $\theta \rightarrow \pi-\theta$. The latter reduces the relevant range to $\theta \in [0,\pi/2]$. As depicted in FIG.~{\ref{Fig:AdSTodS}}, this potential interpolates between an AdS solution at $\theta=0$ and a dS one at ${\theta=\pi/2}$, crossing a Minkowski point at $\theta=\pi/4$.

The fermion masses also allow us to identify the underlying gauge group \cite{Dibitetto:2012ia}  and to compute the amount of preserved supersymmetry and mass spectra as a function of $\theta$ (see  FIG.~{\ref{Fig:AdSTodS}}). Supersymmetry gets totally broken for any value of $\theta$. The analysis of stability using the mass formula in ref.~\cite{LeDiffon:2011wt} will be done separately for each of the three regions: AdS, Minkowski and dS regions.

\subsection{The AdS region $\theta\in [0,\frac{\pi}{4})$}

At $\theta=0$ the critical point corresponds (after normalisation of the overall energy scale) to the unstable solution \eqref{newWarner} of the standard $\textrm{SO}(8)$ gauged supergravity. Its normalised scalar spectrum was presented in (\ref{masses_new_Warner}). At 
 \begin{align}
  \theta=\tan^{-1}(\sqrt{5-2\sqrt{6}})\sim 0.098 \pi \ , 
 \end{align} 
the corresponding spectrum was given in (\ref{masses_new_solutions}). The scalar mass spectra continuously evolve (always with unstable scalars) up to $\theta=\pi/6$ where a change of gauge group occurs. At this value the gauge group jumps from $\textrm{SO}(8)$ to $\textrm{SO}(2) \times \textrm{SO}(6) \ltimes T^{12}$ and the scalar masses read
\be
\begin{array}{ccl}
m^2 L^2 &=&   2 \, (3\pm 2 \sqrt{3}) \,\, (\times 1) \,\, , \,\, 3 \pm \sqrt{33}  \,\, (\times 1) \,\, , \,\, 0 \,\, (\times 47)  \\[2mm]
&& 12 \,\, (\times 1)  \,\, , \,\, 8 \,\, (\times 9) \,\, , \,\, -2 \,\, (\times 9) \ , \nonumber
\end{array}
\ee
still containing an instability. Notice the presence of the two masses $3 \pm \sqrt{33}$ which were found to be the asymptotic values for the pair of solutions running away when ${\omega \rightarrow 0}$ in the scalar potential (\ref{s_Fermi}). In the final part of the AdS region, \textit{i.e.} $\pi/6 < \theta < \pi/4$, the gauge group is $\textrm{SO}(6,2)$ and there are scalar instabilities for any value of $\theta$. A novel feature in this part is that some of the masses diverge as long as $\theta \rightarrow \pi/4$. When approaching this value from the left, the scalar masses tend to
\be
\begin{array}{ccl}
m^2 L^2 &=&  0 \,\, (\times 22) \,\, , \,\, 0 \,\, (\times 1)  \,\, , \,\, 0 \,\, (\times 1) \,\, , \,\, -3 \,\, (\times 9) \\[2mm]
&&   \infty \, \,\, (\times 9) \,\, , \,\, \infty  \,\, (\times 12) \,\, , \,\, \infty  \,\, (\times 16)  \ . \nonumber
\end{array}
\ee
These divergences indicate not only another change of gauge group, but something more dramatic: an AdS/dS transition via a Minkowski point.

\subsection{The Minkowski transition point $\theta = \frac{\pi}{4}$}

At the very special value $\theta = \pi/4$, the critical point has $V=0$ and $\textrm{SO}(3,1)^2 \ltimes T^{16}$ as gauge group. The non-normalised scalar masses are
\be
\begin{array}{ccl}
m^2 &=&  16 \,\, (\times 9) \,\, , \,\, 8 \,\, (\times 12)  \,\, , \,\, 6 \,\, (\times 16) \,\, , \,\, 0 \,\, (\times 33) \ , \nonumber
\end{array}
\ee
proving a tachyon-free spectrum with flat directions. Moreover, the absence of abelian factors in the residual group prevents this solution from the results in ref.~\cite{Dall'Agata:2012cp} regarding destabilisation of flat directions due to one-loop quantum effects to apply. This Minkowski critical point can be understood as the slow-roll limit of the dS solutions in ref.~\cite{Dall'Agata:2012sx}. Notice the matching between the number (and multiplicities) of non-vanishing masses here and that of divergent masses later in the dS region when $\pi/4 \leftarrow \theta$ from the right.

\subsection{The dS region $\theta\in (\frac{\pi}{4},\frac{\pi}{2}]$}

All along the dS region the underlying gauge group is $\textrm{SO}(4,4)$. At the limit value $\theta=\pi/2$, we rediscover the new dS solution of the electrically gauged maximal supergravity in ref.~\cite{Dall'Agata:2012sx}. This dS critical point is unstable and has a scalar mass spectrum given by
\be
\begin{array}{ccl}
m^2 L^2 &=&   -3+\sqrt{3}\pm\sqrt{6 (4-\sqrt{3})} \,\, (\times 1)  \\[2mm]
&&\tfrac{3}{2} (\sqrt{3}+3) \,\, (\times 15)  \,\, , \,\, 2 (\sqrt{3}+3) \,\, (\times 10)  \\[2mm]
&&2 (\sqrt{3}+2) \,\, (\times 9) \,\, , \,\, 2 (\sqrt{3}+1) \,\, (\times 9)  \,\, , \,\, 0 \,\, (\times 22)  \\[2mm]
&&\tfrac{3}{2} (-5+\sqrt{3} ) \,\, (\times 1) \,\, , \,\, 4 \sqrt{3} \,\, (\times 1) \ , -2 \sqrt{3} \,\, (\times 1) \nonumber  \ .
\end{array}
\ee
 When approaching the value  $\pi/4 \leftarrow \theta$, some of the masses diverge. The normalised mass spectrum for the scalars this time tends to
\be
\begin{array}{ccl}
m^2 L^2 &=&  0 \,\, (\times 22) \,\, , \,\, 0 \,\, (\times 1)  \,\, , \,\, 0 \,\, (\times 1) \,\, , \,\, 3 \,\, (\times 9) \\[2mm]
&&   \infty \, \,\, (\times 9) \,\, , \,\, \infty  \,\, (\times 12) \,\, , \,\, \infty  \,\, (\times 16)  \ . \nonumber
\end{array}
\ee
Notice the great resemblance with the behaviour of the masses in the AdS region when approaching the critical value ${\theta\rightarrow \pi/4}$ from the left. More concretely, the dilution of some of the tachyonic masses (when not blowing up to $\infty$) in the vicinity of $\theta=\pi/4$. 

However, there is a crucial difference: the non-diluting tachyons in the AdS side with normalised mass ${m^2 L^2=-3}$ happen to flip sign when jumping into the dS side becoming stable directions with $m^2 L^2=3$ (see FIG.~\ref{Fig:AdSTodS}). For this reason, one can only use the singular limit to approach perturbative stability arbitrarily close in the dS region, and not in the AdS region. Including non-perturbative corrections could even lift the arbitrarily light tachyons of dS and flat directions of Minkowski to genuinely stable moduli.

Before moving to the discussion, we want to point out a remarkable feature of the fermionic mass terms in (\ref{A1_config}) and (\ref{A2_config}). They survived after modding out by the $D_{4}$ group in (\ref{D4_subgroup}), which contains a particular $\mathbb{Z}_{2}$ subgroup generated by $S_{0}$. This $\mathbb{Z}_{2}$ was shown to truncate maximal to half-maximal supergravity \cite{Dibitetto:2011eu}, which implies that the $\omega$-dependent scalar potential in (\ref{s_Fermi}) can be obtained as a type II generalised flux compactification, see \textit{e.g.}~refs~\cite{Aldazabal:2008zza, Aldazabal:2010ef, Dibitetto:2011gm, Dibitetto:2012ia}. Using the correspondence between fermion masses and type IIB fluxes, we have verified that they correspond to generalised flux backgrounds involving NS-NS and R-R fluxes $\{H_{3}(\theta),F_{3}(\theta)\}$, non-geometric fluxes $\{Q(\theta),P(\theta)\}$ as well as the primed version of these. Furthermore, non-geometric fluxes are present during the whole AdS/Mkw/dS chain of transitions. This goes along the line of transitions between geometries suggested in ref.~\cite{DallAgata:2011aa}.

\section{Discussion}

In this paper we  have investigated the interrelation between electromagnetic duality and triality in the new family of $\textrm{SO}(8)$ maximal supergravities of ref.~\cite{Dall'Agata:2012bb}. We have focused on the $\textrm{SO}(4)$ invariant sector of the theory, which derives its interest from distinct mass spectra and cosmological constants for the different critical points. 

We have shown that there are such solutions for all three inequivalent embeddings $\textrm{SO}(4)_{v,s,c}$ \footnote{For completeness we should mention here that the set of three inequivalent embeddings we have analysed is not exhaustive. Nevertheless it constitutes a closed set in which it is possible to recover the $\pi/4$ periodicity.} inside $\textrm{SO}(8)$ that triality allows for. This includes the novel solution (\ref{location_new_solutions}) of standard maximal supergravity. Moreover, our analysis shows that the periodicity of electromagnetic rotations is consistent with $\pi/4$ once triality is taken into account and the structure of critical points for the three embeddings are combined together. Whether the full theory, including couplings that are not tested by an analysis of the vacuum structure, also respects this periodicity remains to be seen; however, we find it encouraging that the intricate structure of the $\textrm{SO}(4)$ vacua passes this test.

In addition, we have also explored a sequence of gauge group transitions connecting AdS solutions of the $\textrm{SO}(8)$ theory to dS solutions of the $\textrm{SO}(4,4)$ theory with parametrically small tachyon masses. The existence of the latter has recently been pointed out in ref.~\cite{Dall'Agata:2012sx} and connected to an ${\textrm{SO}(3,1)^2 \ltimes T^{16}}$ contracted gauging with a Minkowski vacuum. In this paper, we have extended this Mkw/dS transition to a longer AdS/Mkw/dS chain of transitions along theories with different gauge groups by exploiting the GTTO approach. This mechanism of instabilities amelioration around AdS/Mkw/dS transitions could be generic -- it was previously observed in the context of $\mathcal{N}=1$ supergravity \cite{deCarlos:2009qm,Danielsson:2012by} -- and thus serve as a natural place to look for (meta-)stable dS vacua in maximal supergravity. \\

{\bf Note added:} after posting our paper on the arXiv, a related publication appeared which also discusses the issue of periodicity \cite{deWit:2013ija}.

{\bf Acknowledgements} 

We would like to thank Giuseppe Dibitetto and Oscar Varela for very stimulating discussions. AB and DR gratefully acknowledge support by a VIDI grant from NWO. The work of AG is supported by the Swiss National Science Foundation.

\providecommand{\href}[2]{#2}\begingroup\raggedright\endgroup

\end{document}